# Satellite Signal Detection via Rydberg-Atom Receiver

MINGWEI LEI,[1] JIANQUAN ZHANG,[1] QUN LUO,[2] ZHENTAO ZHANG,[2] MING WANG,[2] AND MENG SHI[1, *]

[1] *Key Laboratory of Space Utilization, Technology and Engineering Center for Space Utilization, Chinese Academy of Sciences, Beijing 100094, China*

[2] *National Key Laboratory on Blind Signal Processing, Chengdu 610041, China*

*shimeng@csu.ac.cn*

**Abstract:** Rydberg-atom receivers aim for ultra-high sensitivity to microwave fields through various techniques, but receiving satellite signals has remained a significant challenge, due to the difficulty of capturing weak microwaves over long distances. In this work, we introduce a high-gain antenna to focus satellite signals, and then apply into an atomic cell via a microwave cavity. Using microwave-enhanced coupling, the minimum detectable power of incident microwave is down to -128 dBm, and the corresponding sensitivity is estimated as 21 nV/cm/Hz$^{1/2}$ at frequency of 3.80 GHz. Furthermore, beacon signal from geostationary satellites is captured with Rydberg sensors for the first time, without the need for a low-noise amplifier. And C-band modulated signals are read out with a signal-to-noise ratio of 8 dB. Our results mark a significant breakthrough in facilitating satellite communications using Rydberg-atom receivers.

## 1. Introduction

Rydberg atoms, with their high polarizability and significant electric dipole moments, have the potential to measure external microwave (MW) fields with ultra-high sensitivity, reaching levels as low as -220 dBm/Hz [1] and operating across a broad frequency range from DC to THz [2-5]. Due to these unique properties, Rydberg-atom microwave receivers have been successfully applied in a variety of fields, including wireless communication [6-8], weak-field metrology [9,10], and remote sensing [11]. These applications involve the reception of microwave signals that are amplitude-modulated [12], frequency-modulated [13], or phase-modulated [14].

With recent advancements in microwave sensing, the most concerned issue is to enhance the sensitivity of Rydberg-atom receivers in order to surpass the classical sensitivity limit of -174 dBm/Hz. Several microwave sensing techniques have been proposed, including the superheterodyne technique with a field sensitivity of 55 nV/cm/Hz$^{1/2}$ [15], the many-body effect method with a sensitivity of 49 nV/cm/Hz$^{1/2}$ [16], and the laser repumping method with a noise-equivalent field of 30 nV/cm/Hz$^{1/2}$ [17]. In addition, 2×2 laser array scheme has been employed to further enhance sensitivity to 19 nV/cm/Hz$^{1/2}$ [18], while microwave cavity scheme has demonstrated an increase in sensitivity to 15.8 nV/cm/Hz$^{1/2}$ [19, 20]. To approach the quantum limit, methods such as cold atom techniques [21] and N-wave mixing [22] are also proposed, achieving sensitivities of 10 nV/cm/Hz$^{1/2}$ and 4.0 nV/cm/Hz$^{1/2}$, respectively.

Significant advancements have been made in Rydberg-atom receivers to achieve ultra-high sensitivity to microwave fields. However, most experiments have primarily focused on measuring MW signals from signal generators. In real application scenarios, microwave signals are often transmitted from distant targets such as satellites, which presents a significant challenge to atomic receivers for microwave detection. This is primarily due to the small receiving area of Rydberg atoms, limiting their ability to efficiently capture weak microwave signals over long distances. Recently, one group successfully received S-band microwave signals from the MX satellite using dual-microwave Rydberg spectroscopy [23]. Another group has demonstrated the ability to extract modulated signals from noise to enable remote sensing of soil moisture using Rydberg atoms and satellite signals [24]. Both groups utilized low-noise amplifier (LNA) to assist in receiving satellite signals, with which Rydberg-atom receivers would compete in the future [25].

In this paper, we present the detection of satellite signals using our self-made portable Rydberg-atom receiver. To enhance the microwave field, satellite signals are first focused with a 16-meter parabolic antenna and then loaded into the atomic receiver via a microwave cavity. Through microwave-enhanced coupling, beacon signals from a geostationary (GEO) satellite are captured by the Rydberg sensors for the first time without the need for a low-noise amplifier. Additionally, C-band modulated signals are successfully read out with a signal-to-noise ratio of 8 dB. Our approach marks an important step toward enabling satellite communications using Rydberg-atom receivers.

## 2. Methods

### 2.1 Theory for microwave-enhanced coupling

To enhance the MW-atom coupling, the microwave cavity is applied in our atomic receiver. As the atomic cell allows MW to pass through with minimal absorption, the incident microwave undergoes multiple reflections by the cavity wall forming a standing wave field in the TE$_{101}$ mode, which ensures the efficient interactions between Rydberg atoms and microwave. When incident MW is entirely fed into the cavity with a perfect matching, its circulated power is estimated by

$$P_c = QP_{in}, \tag{1}$$

herein Q represents the quality factor of MW cavity, which is defined as the ratio of the stored MW energy to the energy lost per cycle, and P$_{in}$ denotes the power of incident microwave. The MW cavity with high Q-factor, acts as a passive amplifier to the microwave, which would improve the sensitivity for satellite signals.

### 2.2 Theory for sensing microwave signals

Our approach to sense satellite signals is to measure the probe laser transmission after passing through atomic cell using super-heterodyning technique. For a ladder four-level system, the Hamiltonian is [26]:

$$H = -\hbar\Delta_p \sigma_{22} - \hbar(\Delta_p + \Delta_c)\sigma_{33} - \hbar(\Delta_p + \Delta_c + \Delta_{MW})\sigma_{44} + \hbar/2\,\Omega_p(\sigma_{12} + \sigma_{21}) + \hbar/2\,\Omega_c(\sigma_{23} + \sigma_{32}) + \hbar/2\,\Omega_{MW}(\sigma_{34} + \sigma_{43}), \tag{2}$$

where $\Delta_p$, $\Delta_c$ and $\Delta_{MW}$ are the detuning of the probe, coupling laser and MW field, respectively, and $\Omega_p$, $\Omega_c$ and $\Omega_{MW}$ are the corresponding Rabi frequencies, $\sigma_{ij}$ $(i,j = 1,2,3,4) = |i><j|$ is the transition operators between atomic energy levels $i$ and $j$, and $\hbar$ is the Planck constant. Using the semi-classical methods, the population distribution $\rho$ can be described with the master equation as [27]

$$d\rho/dt = -i/\hbar\,[H,\rho] + \mathcal{L}, \tag{3}$$

where $\mathcal{L}$ is the Lindblad operator for the atoms decay processes [28]. In experiments, we use the on-resonant probe and couple lasers ($\Omega_p = \Omega_c = 0$) and the near-resonant MW fields. For near-resonant case, the probe transmission is a function of the magnitude of MW fields, which can be written as [13]

$$T_{probe} \propto E_{tot} = E_{loc} + E_{sig}\cos(\Delta\omega t + \delta), \tag{4}$$

herein E$_{tot}$, E$_{loc}$, E$_{sig}$ is the strength of the total, local and signal MW field, $\Delta\omega$, $\delta$ is the frequency and phase difference between the local and signal fields, respectively. Via super-heterodyning, Rydberg atoms behave like a MW mixer, which enables down-conversion of C-band satellite signals to low frequency and expands the spectrometer's frequency range.

## 2.3 *Experiment setup*

The setup for the Rydberg-atom receiver is shown Fig. 1(a). A parabolic antenna with size of 16 meters is used to focus C-band MW fields from GEO satellite. The received signal is loaded into a metal microwave cavity by port-coupled methods. Since our experiments are focused on 3.80-GHz satellite radios, the MW cavity is designed as a metal rectangular resonator in the $TE_{101}$ mode, as depicted in Fig. 1(c). The Cs vapor cell, with length of 10cm and diameter of 2cm, is located at the center of MW cavity, where the field strength is the maximum. The 4-level configurations of Cs atoms are realized by $6S_{1/2}$-$6P_{3/2}$-$57D_{5/2}$-$58P_{3/2}$ and presented in Fig. 1(d). The two lowest states are $6S_{1/2}$ (F=4) and $6P_{3/2}$ (F'=4,5), driven by probe laser with power of 50 μW and then excited to $57D_{5/2}$ by coupling laser with power of 20 mW. The probe laser and coupling laser coincide in the atomic cell and propagate in the opposite direction to reduce the Doppler broadening effect. The frequency of the probe laser and coupling laser is, respectively, 852.357nm and 509.236nm, which are locked with an ultra-stable cavity (PDH) to reduce the noise from laser frequency. The detected signals are measured using a photodetector with a bandwidth of 1 MHz, and the EIT spectra are connected to a spectrum analyzer. A local microwave source generates the local microwave, which is used to drive the Rydberg atoms to the $58P_{3/2}$ state via super-heterodyning with the received satellite signals. The resonance frequency between the Rydberg energy levels of $57D_{5/2}$ and $58P_{3/2}$ is 3.778 GHz, which is off-resonant by $\Delta f \approx 22$ MHz with the received satellite signal. The strength of the local microwave is optimized to detune Rydberg energy level to be resonant with the microwave signal. For beacon signals, the measurement is performed without the use of LNA, filter, or RF mixer. For modulated signals with a wider bandwidth of over 10 kHz, the detection is enhanced by adding an LNA to amplify the microwave signal.

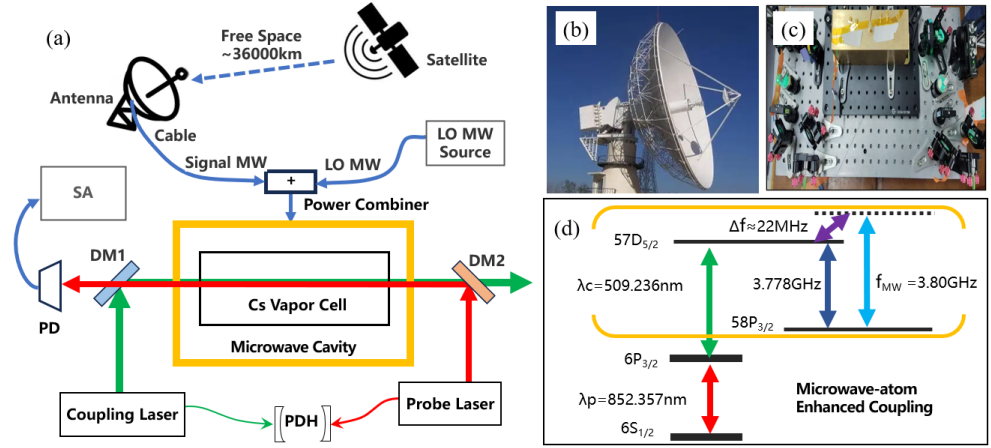

Fig. 1. (a) Schematic diagram of Rydberg-atom receiver for GEO satellite signals, including 852-nm probe laser, 509-nm coupling laser, ultra-stable cavity (PDH), parabolic antenna, local microwave source, power combiner, Cs vapor cell inside microwave cavity, photodetector (PD), spectrum analyzer (SA) and optical components such as dichroic mirror (DM) and reflecting mirror (HR). Photograph of (b) parabolic antenna and (c) microwave cavity used in our experiments are also attached. (d) Relevant energy levels in cesium Rydberg EIT for the C-band microwave measurement.

## 3. Results and discussion

Multiple experiments are conducted to estimate the performance of the atomic receiver before the reception of satellite signal. The sensitivity of the C-band microwave measurement based on Rydberg atoms should be exploited. The signal MW field with frequency of 3.80 GHz is provided by a radio-frequency (RF) signal generator in continuous wave mode, and introduced to the atomic cell inside MW cavity. The EIT spectra for Rydberg state $57D_{5/2}$ are captured with or without MW field by scanning the coupling laser. Presented in Fig. 2(a), the black curve displays the field-free EIT spectrum, used for a reference EIT and the detuning axis is calibrated by the frequency difference between $57D_{5/2}$ and $57D_{3/2}$ of 370.9 MHz. When applying a resonant field into Rydberg atoms, the EIT spectrum exhibits broadening and splitting into two peaks with stronger MW power. As the MW field is off-resonant by $\Delta f \approx 22$ MHz away from Rydberg transitions, the EIT splitting seems asymmetric. In the case of a near-resonant field within the Autler–Townes (AT) regime, the strength of MW field sensed by Rydberg atoms is determined as [1]

$$E_{MW} = -2\pi\hbar\Delta f/\mu, \qquad (5)$$

herein $\Delta f$ represents the spectral splitting in frequency domain, and μ is the transition dipole moment between two Rydberg states. The strength of MW field is measured by AT splitting of Rydberg-EIT peaks at the strong region and shown with red dots in Fig. 2(b). To measure the weak signal MW, a strong local MW with a slight detuning is also loaded to drive Rydberg atoms via superheterodyne, and the EIT spectrum with modulations is further analyzed with a spectrometer. The power of local microwave is optimized to a value of -27dBm, in order to improve the sensitivity for the signal field as high as possible. As shown with the blue dots in Fig. 2(b), the performance of Rydberg-atom system exhibits a linear response to weak microwaves, when the MW power falls below -25 dBm. as Estimated from the noise floor, the minimum detectable power of incident microwave is -128 dBm, and the linear dynamic range extends up to 103 dB. Considering the insert loss of power combiner and extra cables as well as the coupling strength of microwave and Rydberg atoms, the MW strength applied to the atomic cell is unequal to the RF signal generator. By calibrating using the AT splitting, the calculated result is plotted as a red solid line in Fig. 3 (b), derived from linear fitting where the square of the field strength is proportional to the incident microwave power. The sensitivity of electric field measurements by Rydberg-atom sensor is estimated as

$$S = E_{min}/\sqrt{RBW}, \qquad (6)$$

where $E_{min}$ is the detectable field strength, which is measured to be 21 nV/cm, and RBW is the resolution bandwidth of 1 Hz. The strength sensitivity of the Rydberg-atom receiver for C-band microwaves reaches 21 nV/cm/Hz$^{1/2}$ at the frequency of 3.80 GHz, enabling ultra-sensitive measurements of satellite signals.

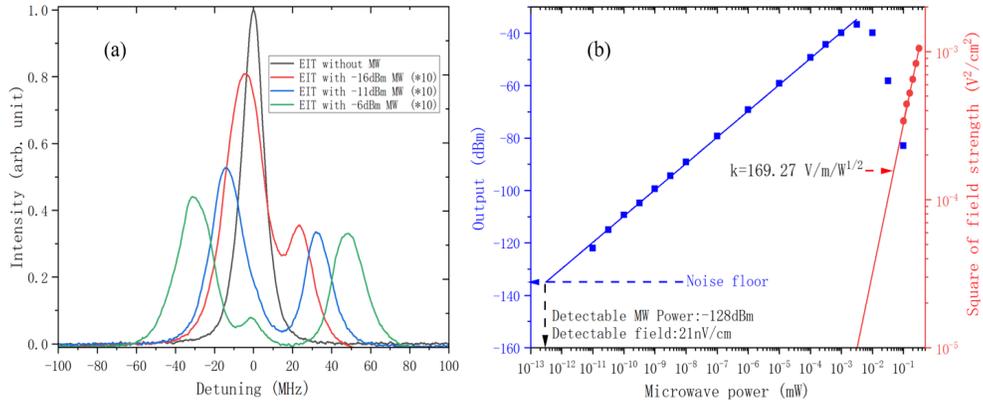

Fig. 2. (a) EIT spectra in the absence (black line) and presence of microwave with different powers (red, blue and green lines, respectively, represent the MW power as -16dBm, -11dBm and -6dBm, and their EIT intensity is amplified by a factor of 10) where the frequency of the coupling laser is scanned across the Rydberg state resonance. (b) Measurement of the weak signal field using the resonant EIT-AT splitting technique. The data are taken with the resonant heterodyne technique (blue dots) and EIT-AT splitting in the strong field regime (red dots). And the blue and black lines show, respectively, the noise floor of atom receiver, the detectable power of incident microwave (-128 dBm). The detectable strength of electric field sensed by Rydberg atoms is 21nV/cm, according to measurements of the MW field strength (red circles) by EIT-AT spectra versus the square root of the incident power and its linear fit (red line) by formula $E = k\sqrt{P}$ with parameter determined as $k = 169.27\ V/m/W^{1/2}$. The results are measured with a resolution bandwidth of 1 Hz.

Further, the signal power transmitted from the satellite to our receiver is also estimated. The free-space path loss L of satellite signals is calculated as [29]

$$L = -32.5 - 20log_{10}^{f} - 20log_{10}^{d}, \tag{7}$$

where f represents the frequency of signal MW as 3.8GHz, and d represents the distance from the transmitter to the receiver in km. For the GEO satellite signals, the distance transmitted to the ground is about 36000 km, and the corresponding pass loss is -195 dB. As single-frequency microwave, the beacon signal from the satellite is firstly investigated, which is used for navigation, monitoring and calibration. To simplify, the transmitter power of beacon signal is at level of tens of watt, which is assumed as ~47 dBm (namely 50W), and the signal power reaching the ground would be -148 dBm. Since the beacon signal is monofrequency and can be measured with a resolution bandwidth of 1Hz, there remains a significant 20 dB difference from the minimum detectable power of the Rydberg-atom receiver. The receiving antenna plays a crucial role in the microwave enhancement before satellite detection, and a high gain receiving antenna is chosen, which is a parabolic antenna with diameter of 16m. And the posture of receiving antenna is optimized to direct towards to the GEO satellite to maximize the signal intensity. The gain of parabolic antenna is estimated as [30]

$$G_A = e_A(\pi d/\lambda_{MW})^2, \tag{8}$$

where $e_A$ is the aperture efficiency of ~0.7, $d$ is the antenna diameter of 16-m, and $\lambda_{MW}$ is the microwave wavelength of 7.88cm. The gain of parabolic antenna is calculated to be 54 dB. Taking into account that 3-dB loss from the cable connected to the antenna and 3-dB loss from conversion of circularly-polarized microwaves to linearly polarized, the effective gain for the satellite signals is reduced to 48 dB. Therefore, the power of the beacon signal coupled to the atomic receiver is boosted to -100 dBm, which is sufficiently strong for direct detection using the Rydberg-atom receiver with a signal-to-noise ratio (SNR) of 28 dB. As shown in Fig. 3, the beacon signal reception is achieved with a SNR of 24 dB, which closely aligns with the theoretical predictions. This is the first reported that satellite signal has been detected using Rydberg-atom receiver without the need for active electronic components such as low-noise amplifiers, mixers, or modulators. Since the high-gain antenna is a passive component, the atomic receiver has the potential to overcome the thermal noise limitations of active electronic devices. For comparison, the same signal from the parabolic antenna is also fed into a commercial microwave spectrometer, where the beacon signal was measured with a SNR of 42 dB. The sensitivity of the Rydberg-atom receiver is 18 dB lower than that of the microwave spectrometer. With further advancements in the sensitivity of the atomic receiver, it will become feasible to measure extremely weaker microwave fields.

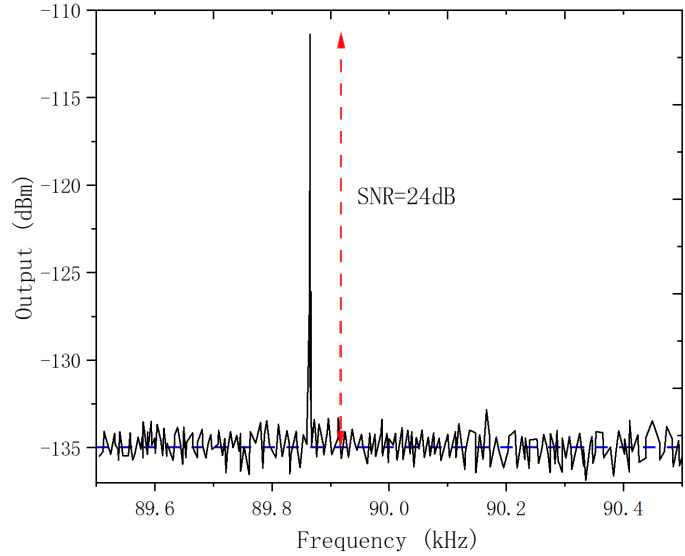

Fig. 3. Beacon signal with a SNR of 24 dB captured by the Rydberg-atom receiver. Herein the blue dashed line shows the noise floor of atomic receiver measured with a resolution bandwidth of 1 Hz.

Finally, modulated microwave signals from ChinaSat satellite are received with the atomic receiver. The center frequency of ChinaSat satellite signals is 3.812 GHz. As the bandwidth of ChinaSat signals exceeds 10 kHz, the required power sensitivity is lower than the capability of our atomic receiver. To further amplify the signal power, a low-noise amplifier with a gain of 60 dB is introduced after the parabolic antenna. Using our Rydberg-atom system, which has a 10 Hz resolution bandwidth, we successfully detected two modulated signals that exhibited a square-wave shape, as shown in Figure 4. These signals had a frequency difference of 400 kHz with the local MW. The narrower signal, with bandwidth of 15 kHz, had a signal-to-noise ratio of ~8 dB, while the wider signal, with bandwidth of 75 kHz, showed a SNR of ~6 dB. Since the SNR of the modulated signals is sufficiently high, it becomes possible to decode the information from the satellite modulated signals captured by our atomic receiver. This marks a significant advancement in satellite signal reception using Rydberg-atom systems and highlights the feasibility of using such receivers for practical satellite communication applications.

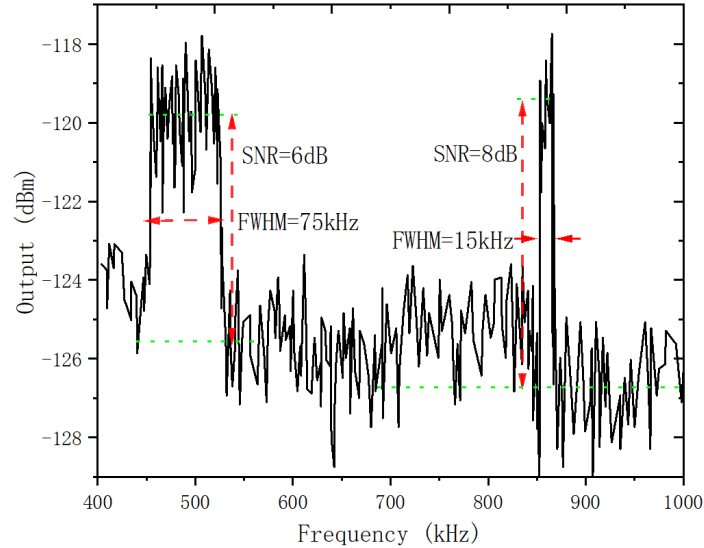

Fig. 4. (a) Modulated signal captured by the Rydberg-atom receiver. Herein the blue dashed line shows the noise floor of atomic receiver measured with a resolution bandwidth of 10 Hz.

## 4. Conclusion

We have successfully demonstrated detections of satellite signals using self-made portable Rydberg-atom receiver. By enhancing coupling between Rydberg atoms and microwave, we have achieved a high sensitivity of 21 nV/cm/Hz$^{1/2}$ at frequency of 3.80 GHz, which is approaching to the traditional MW receiver. Through MW cavity and high-gain antenna, the beacon signal of satellite is detected with high signal-to-noise of 24 dB. It is reported for the first time that satellite signals are able to readout with Rydberg atoms without the need of LNA, mixers or modulator. Furthermore, we explored the reception of satellite modulated signals with signal-to-noise of 8 dB, which is able to decode. Overall, our results mark a significant experimental advancement of Rydberg-atoms sensing for their utilization in a range of satellite applications, such as communication, navigation and metrology.

**Funding.** National Key R&D Program of China (2024YFB3909500); National Science Foundation of China (U2341211); Key Programs of the Chinese Academy of Sciences (RCJJ-145-24-16); Foundation of Technology and Engineering Center for Space Utilization, Chinese Academy of Sciences (CSU-JJKT-2024-4).

**References**

1. H. Fan, S. Kumar, J. Sedlacek, et al., "Electromagnetically Induced Transparency in Atomic Vapor," J. Phys. B: At. Mol. Opt. Phys., 48, 202001 (2015).
2. D. Meyer, Z. Castillo, K. Cox, et al., "Electromagnetically Induced Transparency and Coherent Effects in Rydberg Atoms," J. Phys. B: At. Mol. Opt. Phys., 53, 034001 (2020).
3. Y. Jau and T. Carter, "Coherent Effects and Electromagnetic Transparency in Ultra-Cold Atoms," Phys. Rev. Appl., 13, 054034 (2020).
4. K. Ouyang, Y. Shi, M. Lei, et al., "Advances in Electromagnetic Induced Transparency and Applications," Appl. Phys. Lett., 123, 264001 (2023).
5. S. Chen, D. J. Reed, A. R. MacKellar, et al., "Electromagnetic Transparency and Laser Interactions," Optica, 9(5), 485 (2022).
6. C. Holloway, M. Simons, J. Gordon, et al., "Design and Analysis of Antennas with Electromagnetic Induced Transparency," IEEE Antennas and Wireless Propagation Letters, 18(9), 1853-1857 (2019).

7. D. Anderson, R. Sapiro, G. Raithel, "Electromagnetic Fields and Transparency in Atomic Vapors," IEEE Trans. Antennas Propag., 69, 2455 (2021).
8. Cai, Y., Shi, S., Zhou, Y., et al., "Coherent Electromagnetic Effects in Ultra-Cold Systems," Phys. Rev. Appl., 19(4), 044079 (2023).
9. C. L. Holloway, M. T. Simons, J. A. Gordon, et al., "Electromagnetic Wave Propagation in Transparent Systems," J. Appl. Phys., 121, 233106 (2017).
10. D. A. Anderson, R. E. Sapiro, G. Raithel, "Advances in Antenna Propagation for Electromagnetic Systems," IEEE Trans. Antennas Propag., 69, 5931 (2021).
11. H. Meyer, K. Cox, F. Fatemi, et al., "Electromagnetic Transparency in Controlled Systems," Appl. Phys. Lett., 112, 211108 (2018).
12. D. Anderson, R. Sapiro, G. Raithel, "Electromagnetic Wave Behavior in Atomic Systems," IEEE Trans. Antennas Propag., 69, 2455-2462 (2021).
13. M. Simons, A. Haddab, J. Gordon, et al., "Application of Electromagnetic Induced Transparency," Appl. Phys. Lett., 114, 114101 (2019).
14. C. Holloway, M. Simons, J. Gordon, et al., "Advanced Techniques in Antenna and Propagation Design," IEEE Antennas Wireless Propag. Lett., 18, 1853–1857 (2019).
15. M. Jing, Y. Hu, J. Ma, et al., "Quantum Effects in Electromagnetic Transparency," Nat. Phys., 16(9), 911–915 (2020).
16. D. Ding, Z. Liu, B. Shi, et al., "Enhancement of Electromagnetic Transparency Effects," Nature Physics, 18(12), 1447-1452 (2022).
17. N. Prajapati, A. Robinson, S. Berweger, et al., "Studies on Electromagnetic Transparency in Materials," Appl. Phys. Lett., 119, 214001 (2021).
18. B. Wu, R. Mao, D. Sang, et al., "New Advances in Electromagnetic Systems," IEEE Trans. Antennas Propag., (2024).
19. G. Sandidge, G. Santamaria-Botello, E. Bottomley, et al., "Advances in Microwave Theory and Techniques," IEEE Transactions on Microwave Theory and Techniques, (2024).
20. B. Liu, L. Zhang, Z. Liu, et al., "Quantum Electromagnetic Effects and Applications," arXiv preprint arXiv:2404.06915, (2024).
21. H. Tu, K. Liao, H. Wang, et al., "Quantum Control and Electromagnetic Transparency," Science Advances, 10(51), (2024).
22. S. Borówka, U. Pylypenko, M. Mazelanik, et al., "Electromagnetic Transparency in Photonic Materials," Nature Photonics, 18(1), 32-38 (2024).
23. P. Elgee, J. Hill, K. LeBlanc, et al., "Recent Advances in Electromagnetic Systems and Applications," Appl. Phys. Lett., 123(8) (2023).
24. D. Arumugam, J. Park, B. Feyissa, et al., "Electromagnetic Wave Propagation and Control," Sci. Rep., 14(1), 18025 (2024).
25. D. Shakya, S. Ju, O. Kanhere, et al., "Electromagnetic Innovations in Antenna Propagation," IEEE Transactions on Antennas and Propagation, (2024).
26. C. L. Holloway, M. T. Simons, J. A. Gordon, A. Dienstfrey, D. A. Anderson, G. Raithel, "Electric uncertainties in electromagnetically induced transparency in atomic vapor," J. Appl. Phys., 121(23), 233106 (2017).
27. M. Fleischhauer, A. Imamoglu, J. P. Marangos, "Electromagnetically induced transparency: Optics in coherent media," Rev. Mod. Phys., 77(2), 633–673 (2005).
28. B. Liu, L. Zhang, Z. Liu, et al., "Highly Sensitive Measurement of a Megahertz rf Electric Field with a Rydberg-Atom Sensor," Phys. Rev. Appl., 18, 014045 (2022).
29. T. Rappaport, Wireless Communications: Principles and Practice, Prentice Hall, 2002.
30. C. Balanis, "Antenna Theory and Design," IEEE Transactions on Antennas and Propagation, 17(1), 45-50 (1969).